\documentstyle[11pt,newpasp,twoside,epsf]{article}
\def\edcomment#1{\iffalse\marginpar{\raggedright\sl#1\/}\else\relax\fi}
\reversemarginpar

\begin{document}
\title{HST/NICMOS Imaging of the Planetary Nebula Hubble~12}
\author{Joseph L. Hora  
\affil{Harvard/Smithsonian Center for Astrophysics, 60 Garden St. MS-65,
Cambridge, MA 02138}
\author{William B. Latter}
\affil{ SIRTF Science Center, Caltech, MS 314-6, Pasadena, CA 91125 }
\author{Aditya Dayal}
\affil{ IPAC and JPL, Caltech, MS 100-22, Pasadena, CA 91125 }
\author{John Bieging \& Douglas M. Kelly}
\affil{Steward Observatory, University of Arizona, Tucson, AZ 85721}
\author{A. G. G. M. Tielens}
\affil{Kapteyn Astronomical Institute, PO Box 800, NL-9700 AV Groningen,
The Netherlands}
\author{Susan R. Trammell}
\affil{University of North Carolina, Dept. of Physics, Charlotte, NC 28223}
}
 
\begin{abstract}
Images of Hubble 12 were obtained with the HST/NICMOS instrument 
in the F110W, F164N, and F166N filters of NIC1,
and the F160W, F187N, F190N, F212N, F215N filters of NIC2.
The images show the structure of the inner ``torus" and lobes much clearer than
previous ground-based images.  In particular, the [Fe~II] lobes 
are clearly resolved and shown to be distinct from the H$_2$ emission
structures. 
Apparent changes in the
inner ionized and neutral bipolar shell implies periodic mass loss or
changes in stellar wind shape or direction.
The H$_2$ in the ``eye"
is radiatively excited and shows a complex morphology that suggests that several
mass ejection events are responsible for producing this structure.
The position angle of the  H$_2$ and [Fe~II] lobes differ, 
indicating a possible precession of
the ejection axis. 
\end{abstract}

\section{Observations and Reductions}
Images of Hubble 12 (Hb 12)
were obtained with the HST/NICMOS instrument on 13 Nov 
1997.  The images in the F110W, F164N, and F166N were taken with NIC1,
and the F160W, F187N, F190N, F212N, F215N images were obtained with NIC2.
The MULTIACCUM mode and spiral dither patterns were used.  The  NICMOS
CALNICA and CALNICB pipelines were used to reduce the data.  In general a
few dither sets were obtained in each filter; these were aligned and averaged
to produce the final images.
The bright central star makes the region near the core difficult to see 
clearly, and adds a number of image artifacts.  The lines that run approximately
from corner to corner are part of the diffraction pattern of the instrument.  
The horizontal and vertical lines that run through the star are array artifacts
caused when the central star saturates in the long integrations.  In future
reductions we will attempt to subtract the point source in the core to better
study the region around it.

\begin{figure}
\plottwo{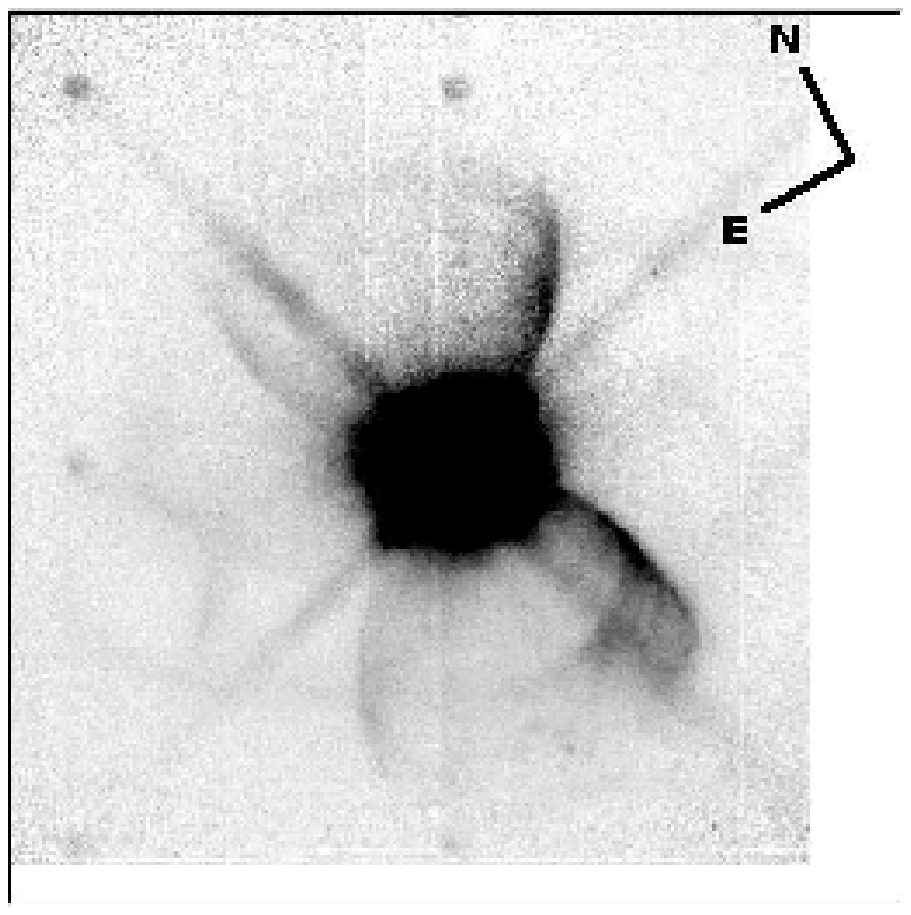}{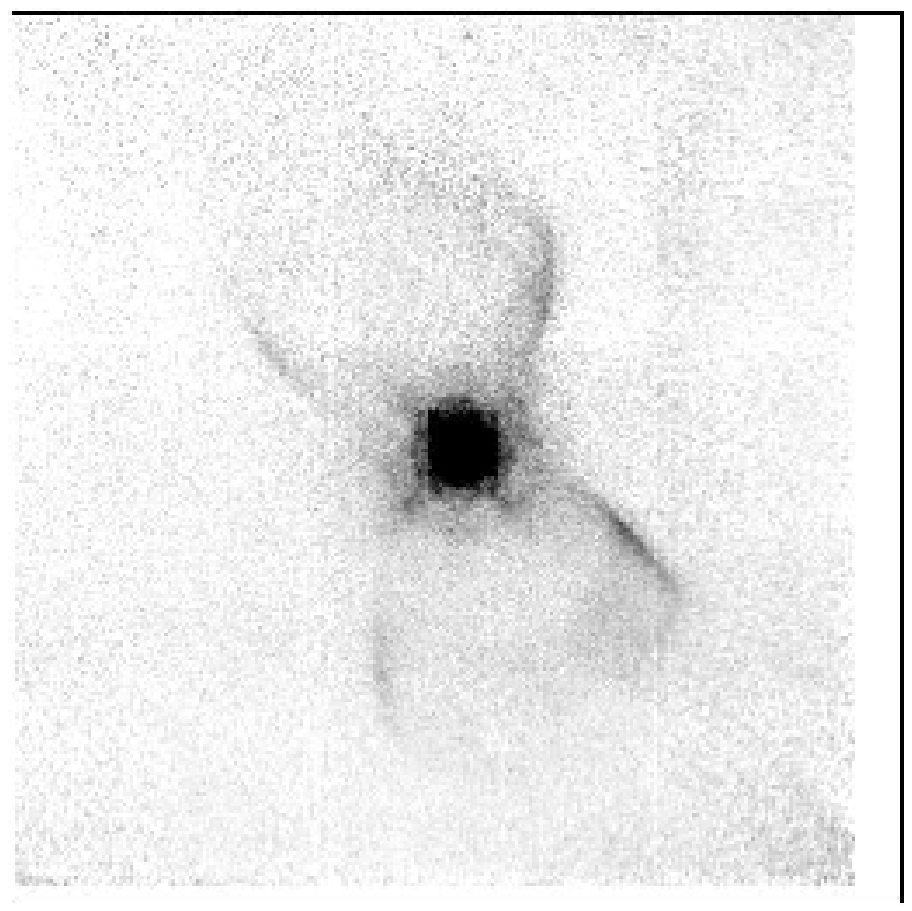}
\caption{Hubble 12, in the F110W (left) and F164N filters (right).  The field
size for each is $\sim$ 13 arcsec square.  The orientation of the images shown 
in the F110W image is the same for all images presented here.  The F110W filter
includes contributions from
the bright Paschen $\beta$ line and line emission from H$_2$, [Fe~II], and He~I
lines within the bandpass. The F164N filter samples the [Fe~II] line
at 1.64 $\mu$m.  }
\end{figure}

\begin{figure}
\plottwo{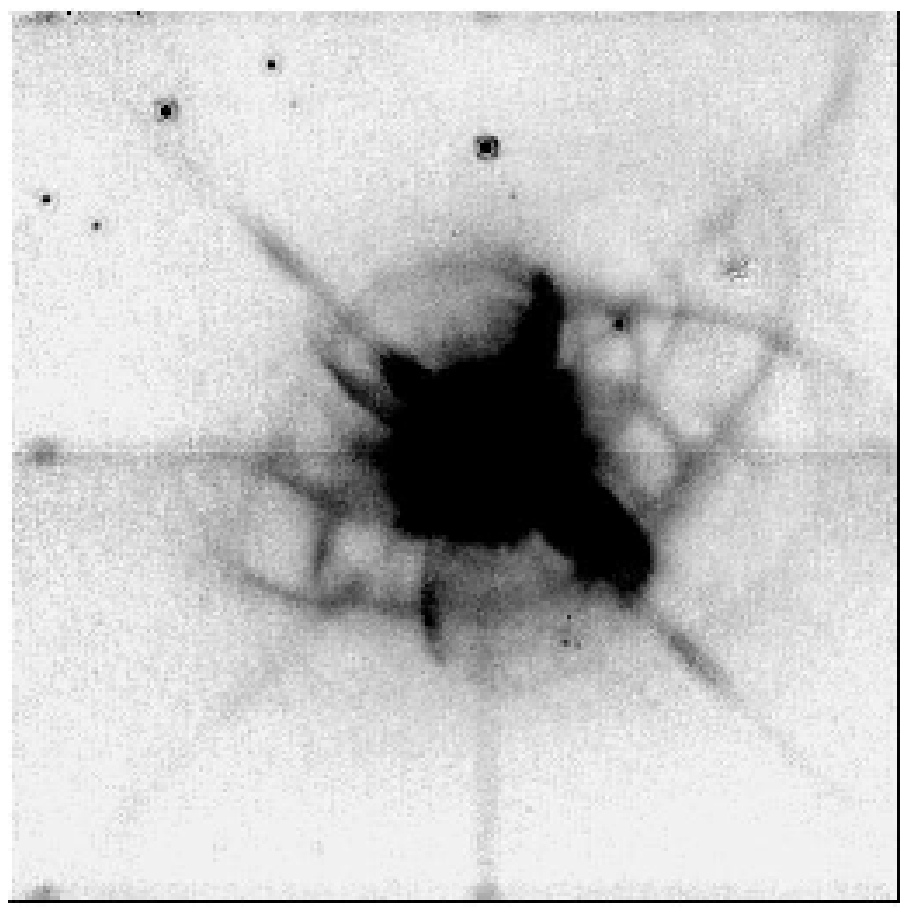}{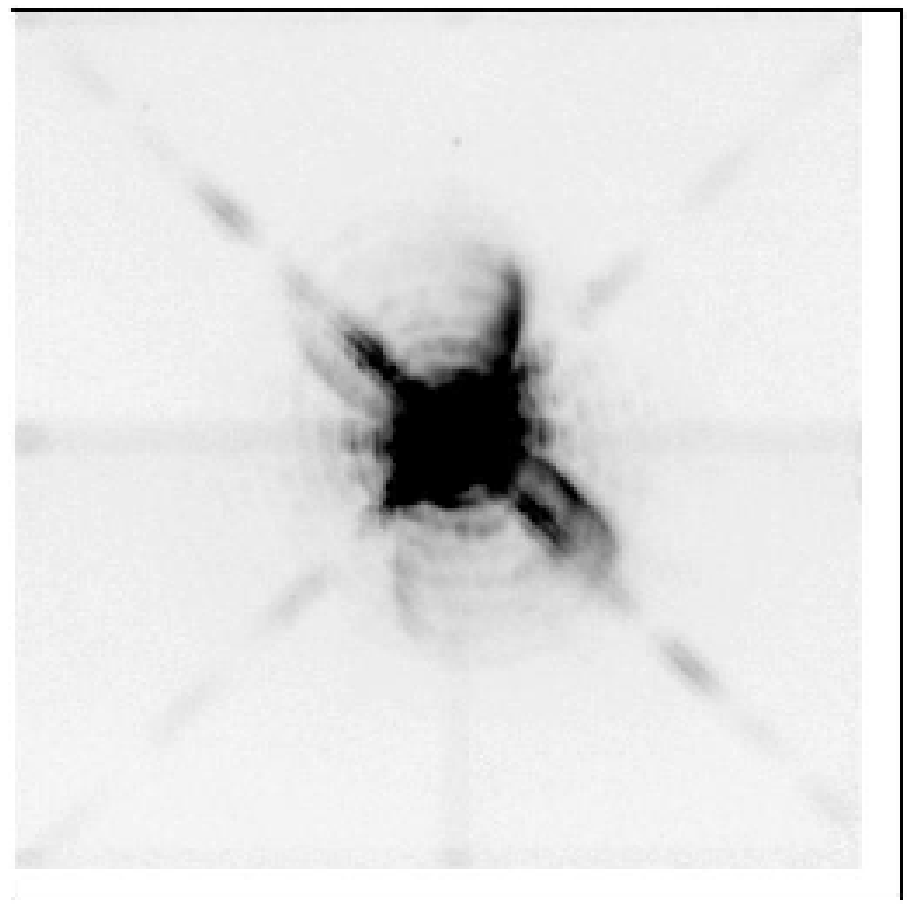}
\caption{Hubble 12, in the F160W (left) and F187N filters (right).  The
field size shown in each image is $\sim$ 19 arcsec square.  
The bandpass of the F160W filter includes lines from the
Brackett series of H~I and line emission from H$_2$, [Fe~II], and He~I lines
as well.  The narrowband F187N filter samples the Paschen $\alpha$ (H~I) line. }
\end{figure}

\section{Results and Discussion}
Hb 12 has been notable primarily because it represents one of
the clearest cases known of UV excited near-IR fluorescent H$_2$ emission
(Dinerstein et al.\ 1988; Ramsay et al.\ 1993, Hora \& Latter 1996, Luhman
\& Rieke 1996).  
Dinerstein et al.\ had mapped the inner structure and found it to be
elliptical surrounding the central star;  the deep H$_2$ images in
Hora \& Latter (1996)  showed the
faint bipolar lobes extending N-S, and the torus or ``eye"-shaped strucure
at the base of the lobes around the central star.  
 The H$_2$ line ratios observed in the torus were in excellent
agreement with predictions by theoretical H$_2$ fluorescence calculations
(see also Luhman \& Rieke 1996).
Hora \& Latter also detected [Fe~II]
line emission at 1.64 \micron\ in a position along the edge of the shell, but
not at the H$_2$ line peak emission location to the E of the central star.

\begin{figure}
\plottwo{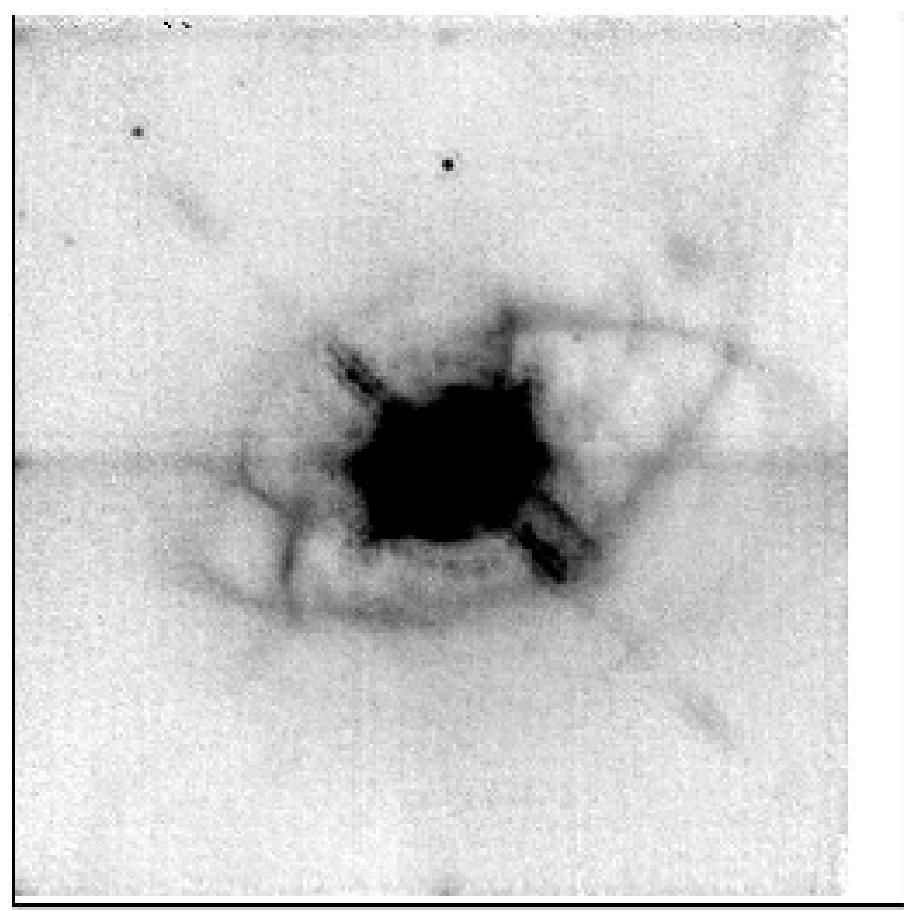}{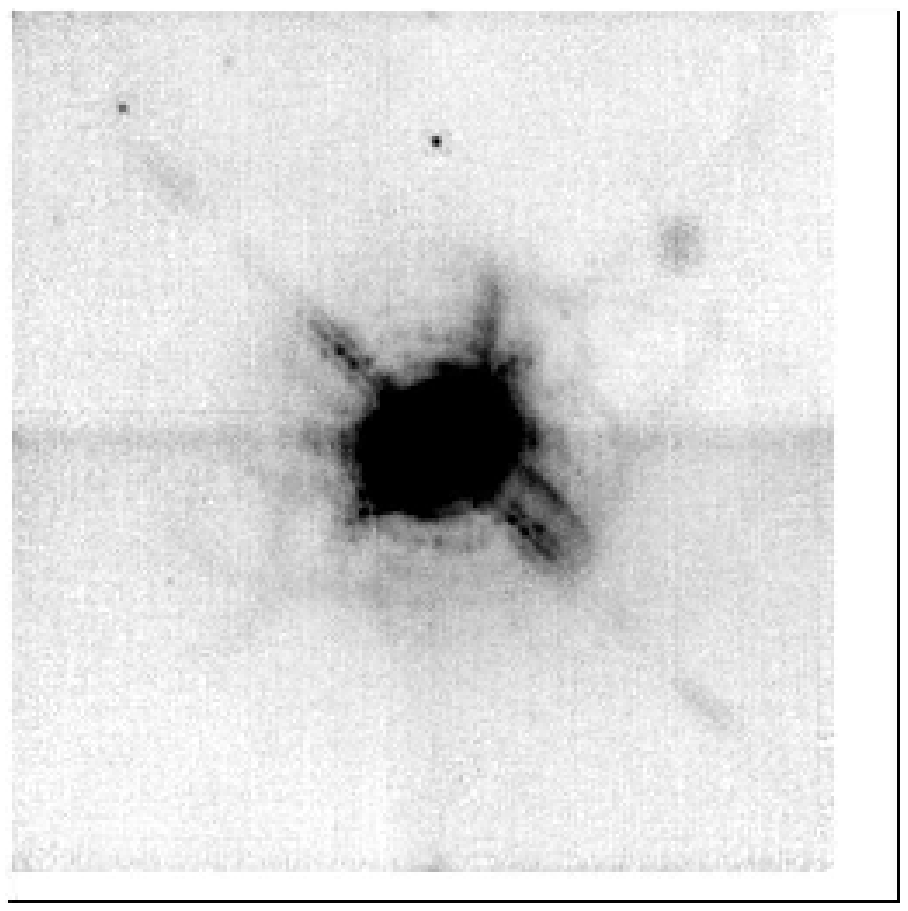}
\caption{Hubble 12, in the F212W (left) and F215N filters (right).  The
field size shown in each image is approximately 19 arcsec square.  
The F212N filter samples the H$_2$ line at 2.12 $\mu$m, the F215N filter
measures the nearby continuum (as well as a small contribution from the 
Brackett $\gamma$ feature at 2.16 $\mu$m.}
\end{figure}

Previous HST-WPC2 imaging by Sahai \& Trauger (1998) in H$\alpha$ showed the inner
structure to have an ``hourglass" shape, and a small bipolar structure in
the core region, with lobes roughly E-W within a few tenths of an arcsec from
the star.  
Welch et al. (1999a,b) obtained ground-based images in the [Fe~II] line and
nearby continuum and found that the line emission was also distributed along
the inner hourglass nebula.   
The HST images presented here show the
symmetry axes of the hourglass and the H$_2$ eye and bipolar nebula 
differ in their alignment
by $\sim$ 5$^\circ$. A comparison of these images with the inner
bipolar structure found by Sahai \& Trauger shows that 
its alignment
differs by $\sim 20^\circ$ from the hourglass and outer H$_2$ lobes.  The 
different orientation of the structures suggests that the central source 
may be precessing between discrete outflow events.  Also, the structures seen 
in the H$_2$ image indicate other possible outflow events and remnants of 
other bipolar hourglass nebulae.  Hb~12 may therefore be another example of
a PN with multiple nested bipolar bubbles.

The inner hourglass is bright in the Paschen $\alpha$ and [Fe~II] lines, but
the outline of the ``eye'' appears only in the H$_2$ and the wide bandpass 
filters (in continuum plus H$_2$ line emission).  This 
implies that the regions where only H$_2$ is detected
are somehow shielded from what has ionized the inner 
hourglass.
There is no evidence  for [Fe~II] emission from other regions in the 
nebula, and the density of the inner hourglass does not seem
sufficient to provide effective shielding of the other regions from radiation
from the central star, which might suggest shock excitation in an interacting
wind.  This hypothesis must be confirmed with investigation of the velocity
structure in the nebula. The strong influence of FUV photons elsewhere in
this object argues that the [Fe~II] emission is excited by FUV photons in the
PDR. We will be investigating this possibility through detailed chemical
modeling.

We obtained high-resolution IR spectra of Hb~12 in the H$_2$ and [Fe~II] lines
using CSHELL at the IRTF (Kelly, Hora, \&
Latter 1999) which indicates that the N lobe is inclined towards us.  We also
have additional medium-resolution spectra of the faint extended H$_2$ lobes to 
determine the excitation properties of the outer nebula.  We
are in the process of analyzing these data along with the optical and
IR imagery to understand the structure of this interesting and complex nebula.

\end{document}